\documentclass[preprint]{vgtc}               

\ifpdf
  \pdfoutput=1\relax                   
  \usepackage{graphicx}                
  \DeclareGraphicsExtensions{.pdf,.png,.jpg,.jpeg} 
\else
  \usepackage{graphicx}                
  \DeclareGraphicsExtensions{.eps}     
\fi%

\graphicspath{{figures/}{pictures/}{images/}{./}} 

\usepackage{microtype}                 
\PassOptionsToPackage{warn}{textcomp}  
\usepackage{textcomp}                  
\usepackage{mathptmx}                  
\usepackage{times}                     
\usepackage{cite}                      
\usepackage{tabu}                      
\usepackage{booktabs}                  

\usepackage{dirtytalk}

\onlineid{1030}

\vgtccategory{Research}

\vgtcinsertpkg

\title{Time-Varying Fuzzy Contour Trees}

\author{Anna-Pia Lohfink\thanks{e-mail: lohfink@cs.uni-kl.de}\\ %
\and Frederike Gartzky\thanks{e-mail: gartzky@rhrk.uni-kl.de}%
\and Florian Wetzels\thanks{e-mail: f\_wetzels13@cs.uni-kl.de}\\ %
\and Luisa Vollmer%
\and Christoph Garth\thanks{e-mail: garth@cs.uni-kl.de}\\ %
\and \parbox{4in}{\scriptsize \centering Scientific Visualization Lab~--~Technische Universität Kaiserslautern}%
}

\abstract{We present a holistic, topology-based visualization technique for spatial time series data based on an adaptation of Fuzzy Contour Trees. 
Common analysis approaches for time dependent scalar fields identify and track specific features. To give a more general overview of the data, we extend Fuzzy Contour Trees, from the visualization and simultaneous analysis of the topology of multiple scalar fields, to time dependent scalar fields. The resulting time-varying Fuzzy Contour Trees allow the comparison of multiple time steps that are not required to be consecutive. We provide specific interaction and navigation possibilities that allow the exploration of individual time steps and time windows in addition to the behavior of the contour trees over all time steps. To achieve this, we reduce an existing alignment to multiple sub-alignments and adapt the Fuzzy Contour Tree-layout to continuously reflect changes and similarities in the sub-alignments. We apply time-varying Fuzzy Contour Trees to different real-world data sets and demonstrate their usefulness.%
} 

\definecolor{awesome}{rgb}{1.0, 0.13, 0.32}

\CCScatlist{
  \CCScatTwelve{Human-centered computing}{Visu\-al\-iza\-tion}{Visu\-al\-iza\-tion techniques}{Treemaps};
  \CCScatTwelve{Human-centered computing}{Visu\-al\-iza\-tion}{Visualization design and evaluation methods}{}
}

\begin{document}


\firstsection{Introduction}

\maketitle

Analyzing time dependent scalar fields is usually done feature based by identifying regions with specific patterns and tracking them between time steps. While these approaches are helpful for specialized tasks, they are not suited to provide a general overview of the complete data set due to their focus on specific features and the chronology. 
We provide a holistic view of the data set and its feature candidates by applying Fuzzy Contour Trees (FCTs, \cite{FCT}), giving a simultaneous overview of the topological structure of the considered time steps. By defining sub-alignments that originate from the \emph{overall alignment} containing all time steps, the development of the field topology can be tracked, while also being able to consider the overall topological structure of the data set and compare the structures of arbitrary subsets, independent of the chronology.

After the introduction of FCTs as a tool for the simultaneous visualization of the topological structure of ensembles, we present time-varying FCTs (T-FCTs), specialized to the analysis of time series data. Since time series data represents a series of individual results, the application of FCTs to this problem is straight forward~--~instead of different ensemble members, we consider different time points. However, application specific tasks require further development of possible user interactions and visualization to ensure usefulness and consistency. After discussing related work and providing background on our previous work in Section~\ref{relatedWork}, we make the following contributions: We present the T-FCT interface in Section~\ref{interface}. Our new contributions in the FCT back-end and front-end are discussed in Sections~\ref{alignment} and~\ref{timeFCT} respectively, with a short overview of the new interaction possibilities in \ref{interaction}. We apply T-FCTs to different real-world data sets in Section~\ref{applications} and conclude with a short discussion on future work in Section~\ref{conclusion}.

\section{Fuzzy Contour Trees and Related Work}\label{relatedWork}
\begin{figure}
    \centering
    \includegraphics[width=0.8\linewidth]{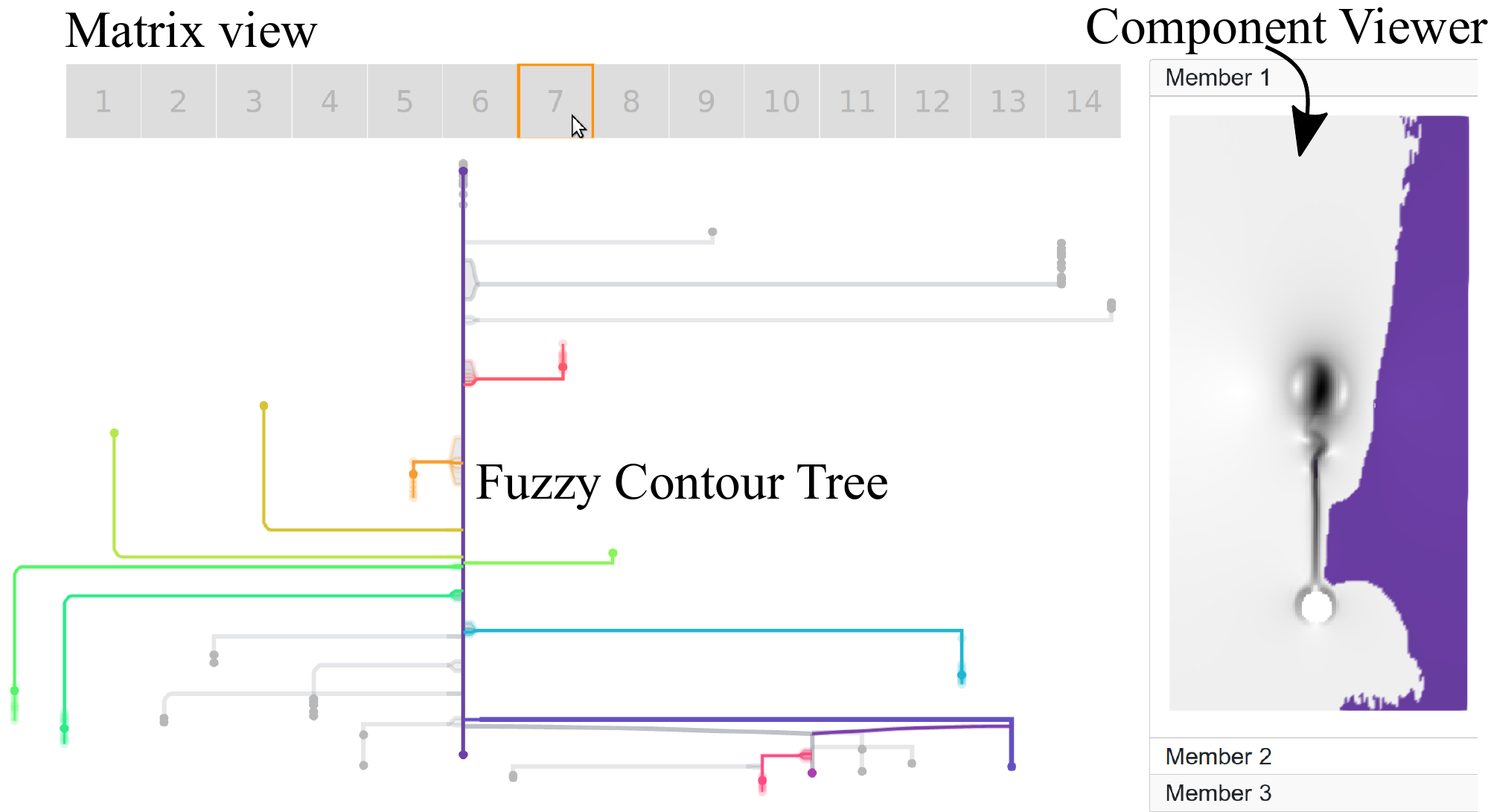}
    \caption{\textbf{The FCT framework} developed by Lohfink et al.}
    \label{fig:oldFCT}
\end{figure}
We give an overview of previous work related with the novel contributions of T-FCTs. Work that is related to FCTs in general, concerning contour trees, their application in ensemble visualization and merging of graph-based topological descriptors can be found in \cite{FCT}.

\paragraph*{Time Dependent Contour Trees}
A theoretical consideration of time-varying reeb graphs for continuous space-time data was given by Edelsbrunner et al. \cite{EDELSBRUNNER2008149}. Szymczak describes sub-domain aware contour trees and uses them to track accumulated topological changes between slices of the data set. While the evolution of iso-surfaces is plotted, the contour trees are not visualized \cite{szymczak}. 

Also different interactive tools for the analysis of time varying contour trees and iso-contours have been developed:
An interactive exploration tool for split/merge and contour trees for different time steps was developed by Sohn et al. \cite{sohn}. They define a topology change graph and use it to navigate trees of individual time steps.
Bajaj et al. provide multiple calculated signature graphs on time-varying scalar fields. Interaction possibilities and the additional visualization of single contour trees provide real-time exact quantification in the visualization of iso-contours \cite{contourSpectrum}. Kettner et al. take this idea further to non-decomposable topological properties and higher dimensions in the Safari interface \cite{KETTNER200397}. Lukasczyk et al. define and visualize spatio-temporal Reeb graphs in \cite{hotspots} to extract and visualize trajectories and relationships of hotspots.
Oesterling et al. show the evolution of extrema in high-dimensional data by plotting a 1D landscape profile for each time step and connecting peaks to indicate critical events. These events are determined as structural changes in time-varying merge trees.

Other than T-FCTs, none of these approaches visualize contour trees for more than a single time step or show the evolution of contour trees.


\paragraph*{Feature Tracking in Time Dependent Data}
Features have been identified and tracked in scalar fields in many different ways. The components of interests can be superlevel sets \cite{hotspots, sohn} or sub-domains with special geometric and topological properties \cite{Bremer_2007, MaddenJulian, schnorri}. Multiple features on different levels of interest are often tracked in tracking graphs \cite{hotspots, sohn, nestedtrackinggraphs, widana}.

In all of these cases, determined features are tracked between subsequent time steps, tracing these features over time. In contrast, our approach provides an overview over topological features instead of tracking individual features. Considering the topological structure at all time steps and matching it, we provide a more holistic view on the data. Similarities and differences of time steps can be determined flexibly and independent of the adjacency of the time steps.



\paragraph*{Fuzzy Contour Trees}
The foundation for T-FCTs was developed by Lohfink et al. with the Fuzzy Contour Tree framework \cite{FCT}. See Figure~\ref{fig:oldFCT} for an overview. This framework allows the simultaneous analysis of the topological structure of all members of an ensemble at once. As underlying back-end functionality, the member contour trees are matched using the tree alignment algorithm. This is done incrementally: after aligning two contour trees, the resulting alignment and a further contour tree are given as input for the alignment algorithm and so on. Since the order of the contour trees has an impact on the result, the contour trees are shuffled randomly before the execution. 

The resulting alignment is a super tree of all contour trees. Thus, a layout for the alignment can be transferred to the individual contour trees, yielding common layouts for all of them. Superimposing the contour trees with this layout results in a Fuzzy Contour Tree in its grouped layout. Bundling edges at saddles and linking the edges' opacity to the fraction of contour trees that contain this specific edge results in the bundled layout. To further improve the clarity of the visualization, Lohfink et al. implemented an optimized branch spacing method that improves the positioning of branches on their parent branch while keeping certain properties of the original FCT. Multiple interaction possibilities are available that link the FCT to an overview of the contained ensemble members. Both, FCTs and T-FCTs were designed to support certain domain-agnostic, elementary analysis tasks. Due to space limitations we reserve their detailed description and evaluation for future work.

\section{Interface Overview}\label{interface}
\begin{figure}
    \centering
    \includegraphics[width=\linewidth]{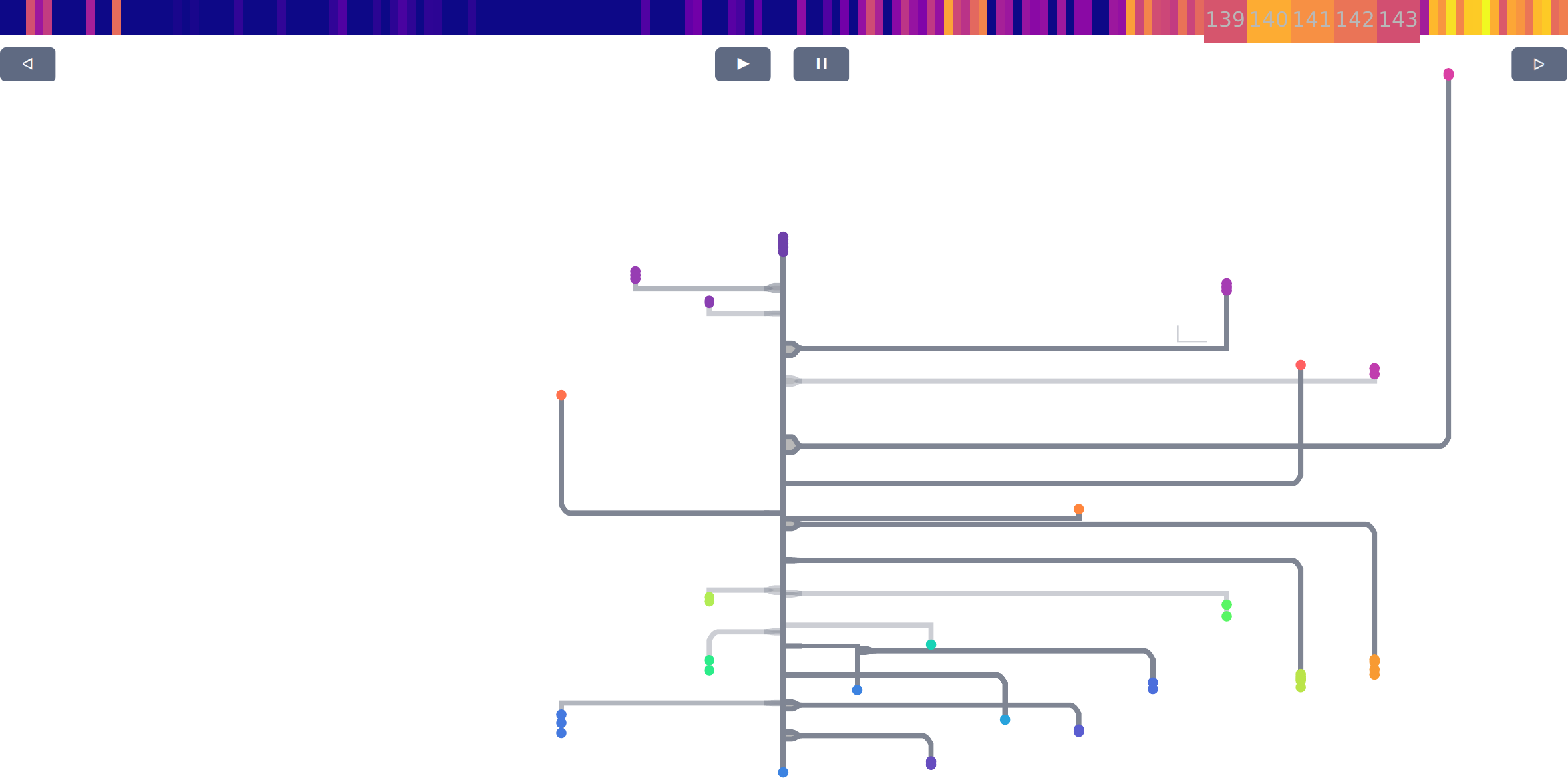}
    \caption{\textbf{The time-varying Fuzzy Contour Tree interface}: Time selector and the FCT of the selected time steps.}
    \label{fig:timeFCTinterface}
\end{figure}

Similar to the original FCT described by Lohfink et al. in \cite{FCT} that is applied to ensembles, the T-FCT gives insight in the topological structure of time dependent data by simultaneous visualization of multiple contour trees. In addition, T-FCTs are tailored to the application to time dependent scalar fields, considering different time steps of the same simulation or measurement. This specialized application gives us the possibility to adapt visualization of and interaction with the FCT to support the exploration of time dependent fields at the best. Figure~\ref{fig:timeFCTinterface} shows the T-FCT interface. Both, the time selector on top and the FCT below are enhanced versions of interface components of the original FCT interface, providing specialized interaction with time dependent data and ensuring a consistent visualization. 

In the following we describe the background of this visualization and how to generate it. Details on the T-FCT layout and the time selector can be found in Sections~\ref{timeFCT} and~\ref{interaction} respectively. The T-FCT interface was implemented using $D^3$, python and jupyter notebook.
\section{The Time-Varying Alignment}\label{alignment}
Considering time dependent scalar fields, the main focus lies on a consistent alignment over time, prohibiting a randomized order of the input contour trees. Instead, the contour trees are aligned sequentially. Since the structure of the first aligned time step imposes its structure on the whole alignment, other choices for this time step might be beneficial depending on the data set. 
Furthermore, the matching of nodes in consecutive time steps is prioritized as follows: the scalar values of nodes in the alignment are set to the value of the nodes in the lastly aligned contour tree, if they exist in this contour tree. Else the value remains untouched. Thus, this contour tree is effectively contained in the alignment with its scalar values. Hence, aligning consecutive time steps behaves similar to matching adjacent trees, enforcing the consistent matching of nodes. Since for the final layout, only values from the individual time steps are used, changing the assigned values in the alignment has no impact on node ordering.

The alignment process is very flexible in its application due to use of different metrics. We implemented an overlap metric in addition to the existing volume, persistence and combined metrics. It is defined as $1-J(A,B)$ where the Jaccard index $J(A,B) = \frac{|A\cap B|}{|A\cup B|}.$ The contour tree alignment algorithm with optional matching over time is available in the TTK development branch \cite{TTK}.

Analysis of time dependent data often takes place on the level of individual time steps instead of the whole data set. Comparing the topological structure of arbitrary sub-sets is rendered possible by the calculation of the corresponding \emph{sub-alignment}.

Sub-alignments are calculated based on an existing overall alignment.
This ensures consistent matching of nodes in different sub-alignments, avoids problems with node identification between sub-alignments and guarantees a consistent layout. To obtain a sub-alignment, contour trees of time steps that are not selected are one by one subtracted from the alignment by decrementing the frequency of all contained nodes. Then, nodes with frequency 0 are deleted and connectivity of the result is restored. To do so, 
we determine parent - child relations of the neighboring nodes by determining paths from these neighbors to the fixed alignment root (see~\cite{FCT} for details) and connect the nodes accordingly.

The resulting sub-alignments are alignments of the chosen sub-set of contour trees. They are likely to be less optimal with respect to the chosen metric than the result of the alignment algorithm heuristic would be. However, all we need sub-alignments to be is a sub-tree of the overall alignment to have a consistent and traceable layout. Employing the resulting FCT, analysis tasks can be approached for the selected sub-set of time steps.
\section{The Time-Varying Fuzzy Contour Tree}\label{timeFCT}
\begin{figure}
    \centering
    \includegraphics[width=\linewidth]{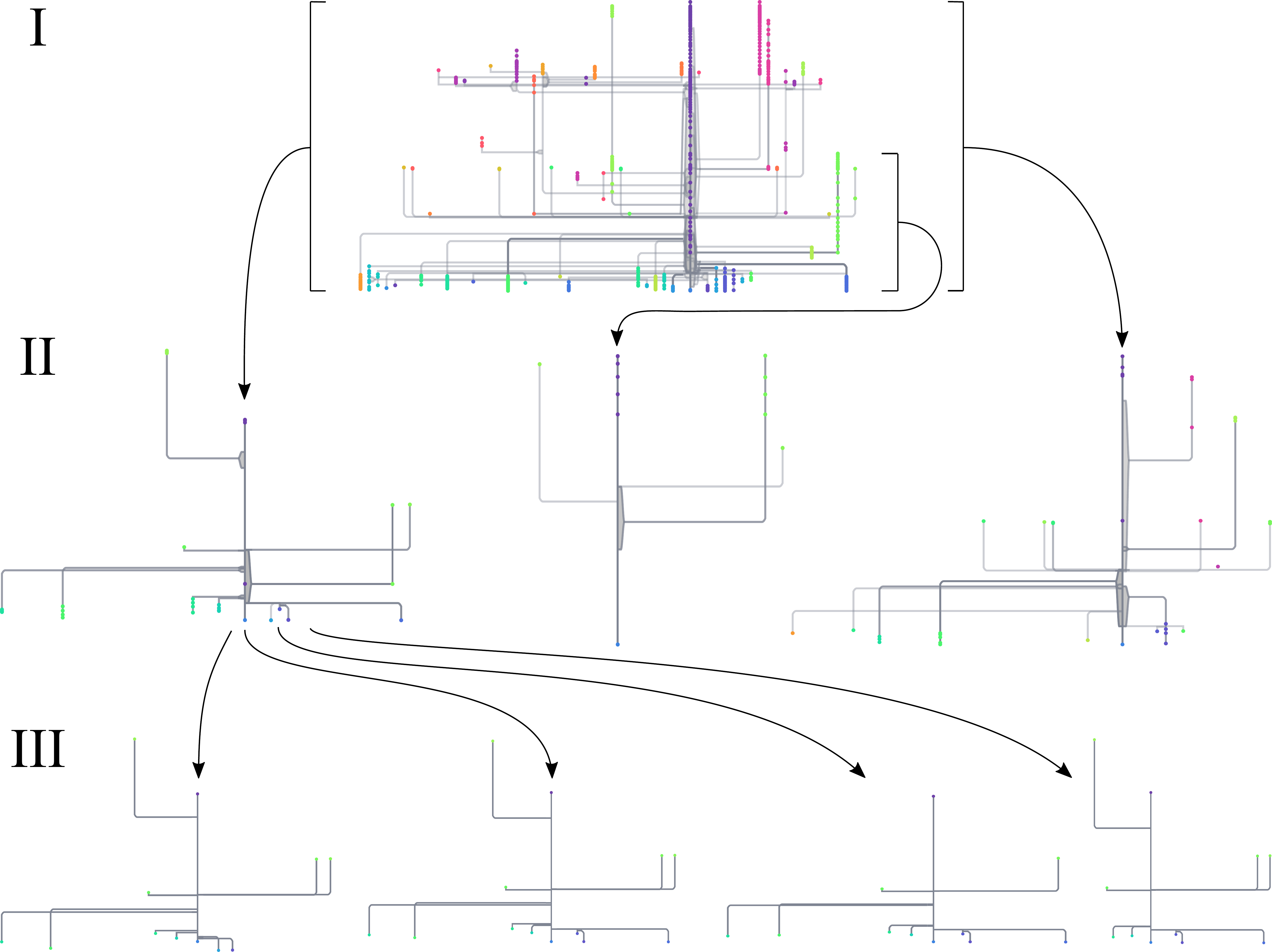}
    \caption{\textbf{The trickle-down-layout of the T-FCT}: The layout of the overall alignment (Level I) carries over to sub-alignments (Level II) that are compared in the time selector. Their layout propagates further to the individual contour trees (Level III).}
    \label{fig:trickle-down-layout}
\end{figure}



As a basis for the T-FCTs, the layout of the overall alignment is computed as described in \cite{FCT}. In the setting of T-FCTs, the cost function of the simulated annealing doesn't only take into account scalar values of leaves and the persistence of branches, but also their existence in time. Only contemporary branches (i.e. branches whose existence intervals overlap) with overlapping bounding boxes are regarded as overlapping. 

Similar to the layout of individual contour trees being based on the layout of their alignment, the sub-alignments' layout is based on the layout of the overall alignment. An illustration for this \say{trickle-down layout} is given in Figure~\ref{fig:trickle-down-layout}. 

To obtain the layout of a sub-alignment from the overall alignment, several steps are taken: First, its branch decomposition is computed. Then, the order of its branches is set according to the horizontal order in the overall alignment. The main branch, as the only branch containing two leaf nodes, is positioned according to its leaf node that can vary over time (the root node always remains the same). This facilitates spotting changes in the main branch.


With the horizontal order of the branches set, it is optionally adapted to avoid gaps that occur if branches from the overall alignment are not present in the sub-alignment. While this results in more uniform individual layouts, tracking the development of individual branches is easier without re-positioning, especially in animations. 


After the layout for the sub-alignment is determined, leaves, saddles and branches are transferred from the individual contour trees as described in \cite{FCT}, following paths in the sub-alignment. The coloring of leaf nodes is kept consistent over different sub-alignments.



In contrast to common feature tracking methods, T-FCTs allow the direct comparison of non-consecutive time steps as well as a topological overview over all time steps. Matching order and metric are adaptable, providing in-depth control of the result. 

\section{Interaction}\label{interaction}
Due to space limitations, we present added and enhanced interaction possibilities only shortly. For more details see the FCT paper \cite{FCT} and the supplemental material. 
\begin{figure}
    \centering
    \includegraphics[width=\linewidth]{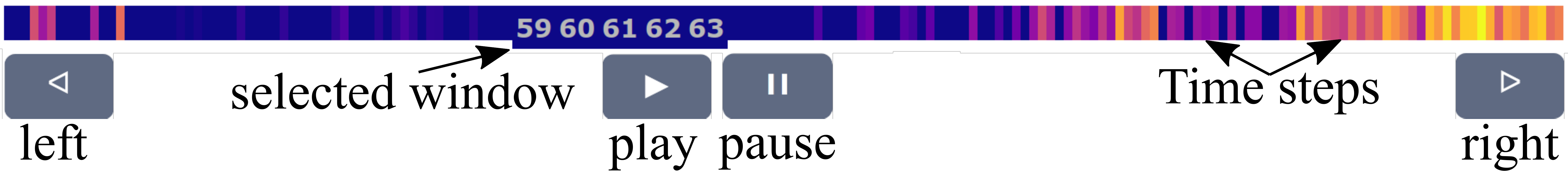}
    \caption{\textbf{The time selector}: The complete data set with highlighted selected time steps. Buttons allow the manipulation of the selection.}
    \label{fig:timeselector}
\end{figure}
Interaction possibilities with the FCT aim on linking information concerning individual contour trees with information on the overall behavior of the individual fields. To adapt FCTs to the time-varying setting, the grid providing information on individual contour trees and enabling tree and member highlighting was turned into the time selector (Figure~\ref{fig:timeselector}, compare Figure~\ref{fig:oldFCT}). Instead of showing only time steps that are contained in the current sub-alignment, every time step is represented by a colored slice of the selector. Selected time steps are highlighted as boxes with the number of the time step. 



\begin{figure}
    \centering
    \includegraphics[width=\linewidth]{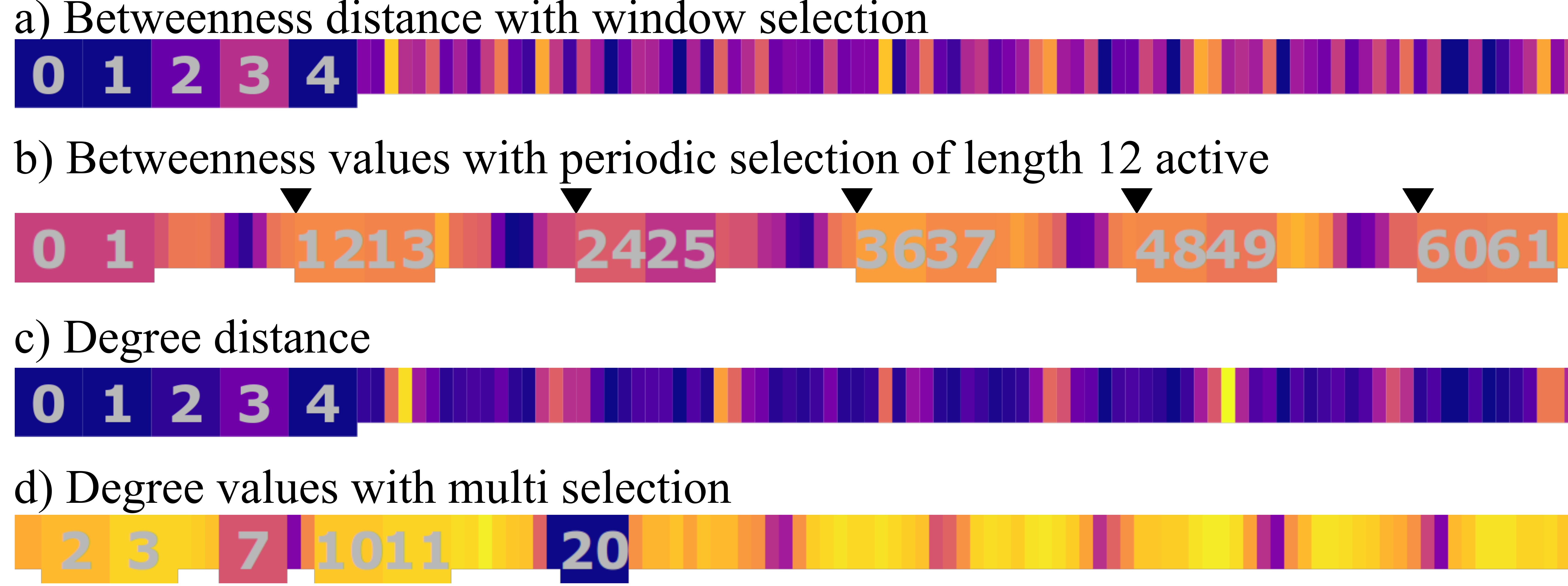}
    \caption{\textbf{Time selector options} Different selections and coloring.}
    \label{fig:time_selector_modes}
\end{figure}

To indicate areas of interest, the time selector is colored based on different measures on the sub-alignment containing a time window of given width with the considered time step at the center. Either these values are shown directly, or the averaged sum of their node-wise differences with the neighboring time step to illustrate changes between them. We implemented two centrality measures: degree and betweenness centrality \cite{roy}, focusing on edge insertions and deletions, and structural changes respectively. Examples for the different coloring options are given in Figure~\ref{fig:time_selector_modes}.

Time steps can be selected via the time selector in three different modes: \emph{Window selection}, \emph{multi-selection} and \emph{periodic selection} as special case of multi-selection. See Figure~\ref{fig:time_selector_modes} for examples. 

\emph{Window selection} selects a connected sub-interval centered at the selected time step. \emph{Multi-selection} allows the manual selection of multiple, not necessarily adjacent time steps to provide deeper insight in individual members and their differences. As a special case of multi-selection, the \emph{periodic selection} automatically selects time steps with a given period and shows a periodic marker.

A given selection can be shifted left and right. A play and a pause button start and stop automated shifting to the right (Figure~\ref{fig:timeselector}).

\begin{figure}
    \centering
    \includegraphics[width=0.8\linewidth]{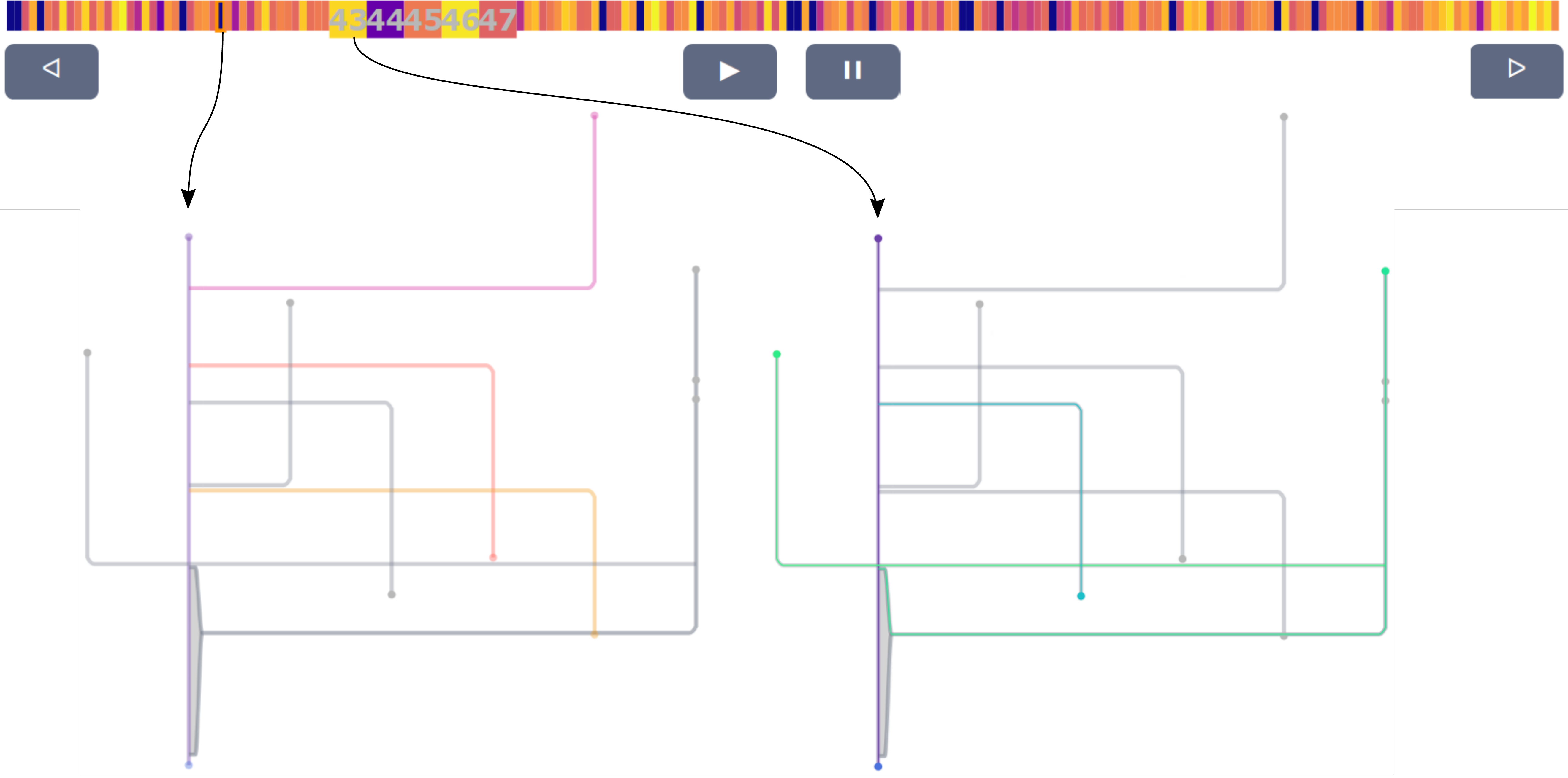}
    \caption{\textbf{Tree highlighting}: Right: Hovering selected time steps highlights the individual contour tree in the FCT. Left: For time steps that are not part of the current sub-alignment, only contained branches are highlighted without indication of specific nodes and saddles.}
    \label{fig:tree_highlighting}
\end{figure}

\emph{Tree highlighting} is also generalized for time steps that are currently not selected: Branches that are present are highlighted without marking specific saddles and leaves (see Figure~\ref{fig:tree_highlighting}).

\begin{figure}
    \centering
    \includegraphics[height=3.4cm]{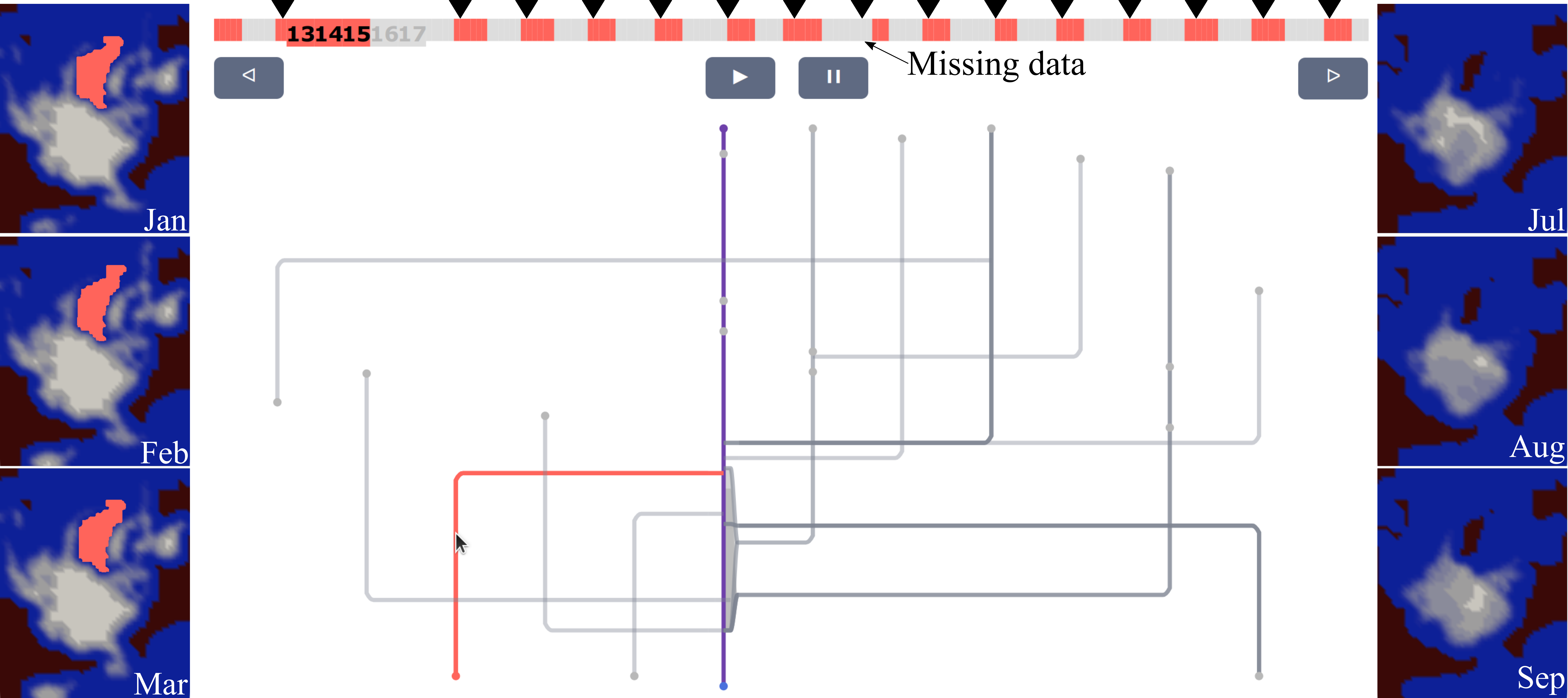}
    \caption{\textbf{Periodic behavior} in the sea ice data set. The marked branch is contained in the contour tree only during winter months. (period 12, volume metric)}
    \label{fig:seaice_periodic}
\end{figure}

\emph{Member highlighting} in the time selector does not only take place for currently selected time steps, but for all available time steps. Like this, navigating the data set and finding patterns in branch occurrences is facilitated. See Figure~\ref{fig:seaice_periodic} for an example. 

\section{Applications}\label{applications}
\subsection{Sea Ice}\label{seaice}
The sea ice data set describes arctic sea ice concentrations and is provided by the National Snow and Ice Data Center \cite{seaice}. The concentrations are given as tenths of grid square area covered by ice. We set the value -1 for grid cells that are not over sea. We applied T-FCTs to sea ice concentrations from 1980 to 1995 and applied a mild smoothing filter to the data to avoid non-binary contour trees. 

Containing only discrete values, large plateaus in the data result in contour trees with branches of persistence 0, forbidding the use of metrics that rely on persistence in the alignment process. Hence, we chose volume and overlap metric for this example. The discrete nature of the saddles and extrema also complicates finding a suitable layout for the T-FCT. Optimized branch spacing as described in \cite{FCT} shifts the branches to still allow a clear layout.

\begin{figure}[t]
    \centering
    \includegraphics[height=3.4cm]{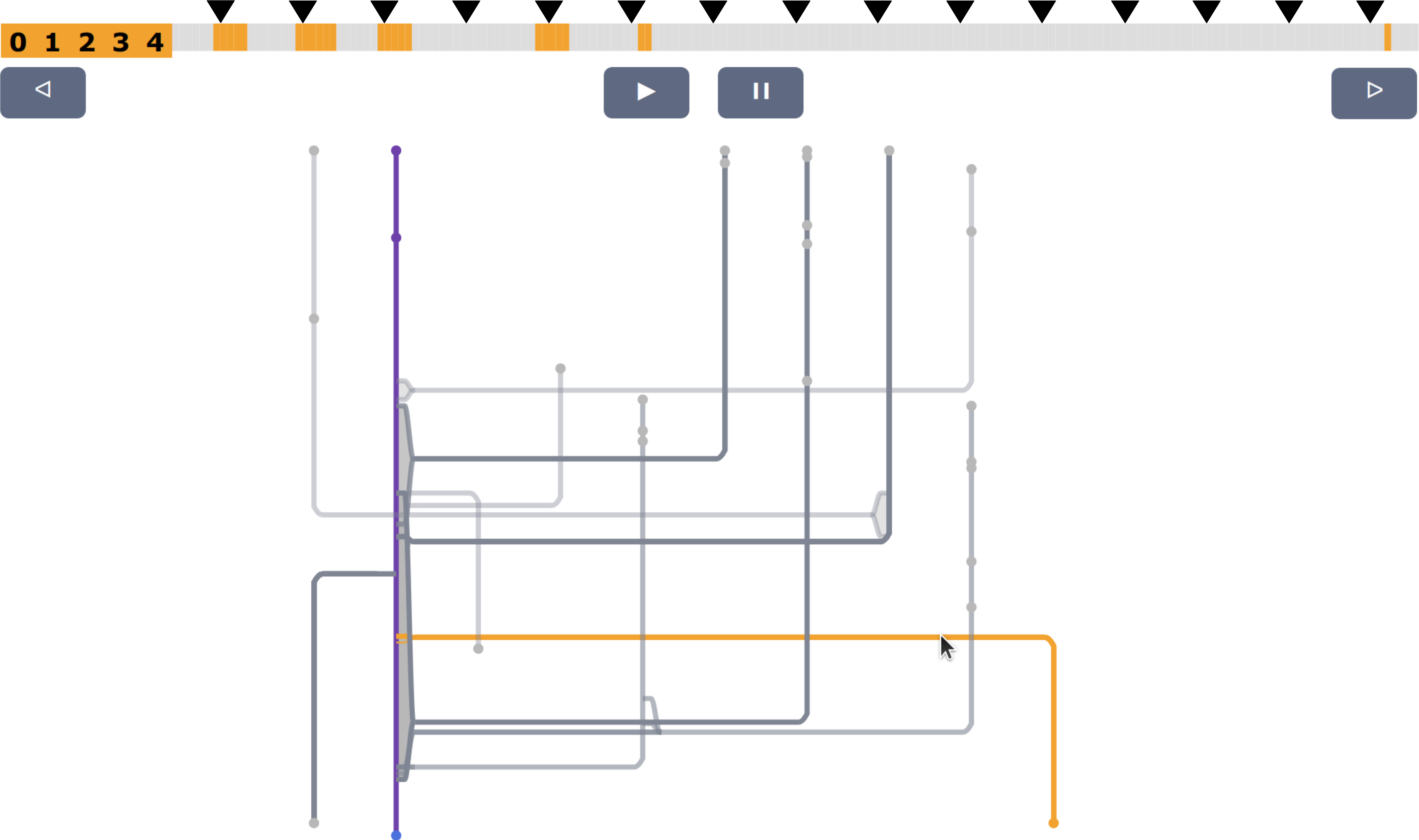}
    \caption{\textbf{Vanishing branches over time} show the yearly decreasing extent of the sea ice. In 1994 the sea ice extent peaked and the branch re-occurs (period 12, volume metric).}
    \label{fig:seaice_vanishing}
\end{figure}

Figures~\ref{fig:seaice_periodic} and~\ref{fig:seaice_vanishing} show the T-FCT for the sea ice data set with branches of typical behavior highlighted. The selected branch in Figure~\ref{fig:seaice_periodic} can be found in the winter months of every selected year, reflecting the periodical increase of sea ice during winter months. The highlighted minimum (ocean) is only matched if it is surrounded by sufficiently large maxima (sea ice). In Figure~\ref{fig:seaice_vanishing} on the other hand, the selected branch vanishes after 1986, to re-occur in early 1994. This branch represents an area of the arctic sea that was frequently surrounded by sea ice during winters before around 1985 but not any more. Both of these behaviors, periodical occurrence over the whole data set and periodical occurrence only in about the first half of the considered time frame with potential re-occurrence between 1992 and 1994 can be seen in multiple branches. 

\begin{figure}[t]
    \centering
    \includegraphics[width=\linewidth]{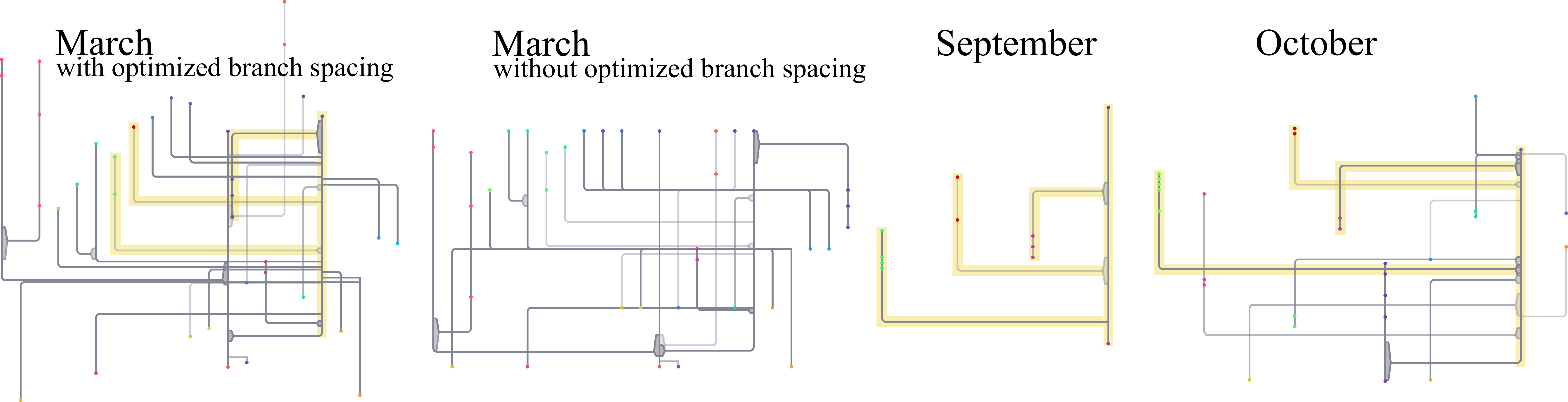}
    \caption{\textbf{Sea ice extent during individual months.} Similar structures are highlighted. For March, the comparison between the FCT with and without optimized branch spacing is given (overlap metric).}
    \label{fig:seaice_months}
\end{figure}

Periodic selection allows the selection of individual months over all years. An example for the comparison of individual months from 1980 to 84 is given in Figure~\ref{fig:seaice_months}. Here, we used the overlap metric for the alignment and illustrate the usefulness of optimized branch spacing in discrete data sets. The much higher complexity of the main structures in the FCT during winter months indicates the larger extent of sea ice in different ice floes. Although these floes are interrupted by islands and potentially even free to move, complicating a matching between time steps, similar structures are visible between the months, indicating areas of the arctic ocean that are covered by ice around the whole year. 


\subsection{Convection Simulation}
\begin{figure}
    \centering
    \includegraphics[width=0.8\linewidth]{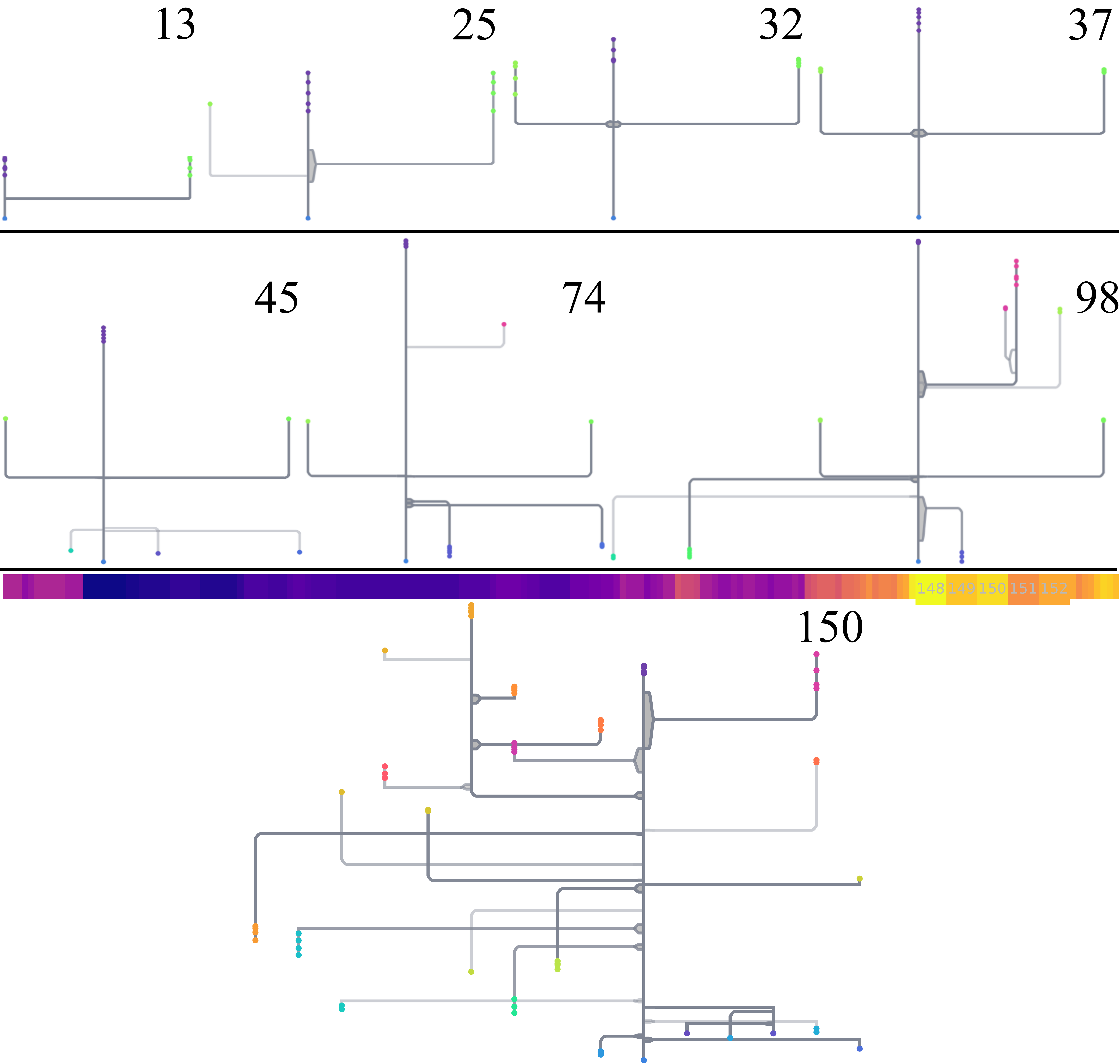}
    \caption{\textbf{Sliding Window}: Advancing step by step through the convection simulation result provides a clear understanding of the ongoing processes. We chose an expressive subset of time steps, the step number is given above the FCTs.}
    \label{fig:heated_cylinder}
\end{figure}

Flow around a hot pole over time was simulated. The resulting 2D flow field is of size 128 × 256. Material at rest is heated around the pole, begins to rise, and forms a plume. An exemplary member is shown on the right of Figure~\ref{fig:oldFCT}. This behavior is clearly reflected in the T-FCTs in Figure~\ref{fig:heated_cylinder}. Advancing the selected window step by step through the data provides an overview of the topological structure at individual time steps but also their connection and development: Until time step 74, the pole is heating up. Then, the plume forms and more and more temperature minima are enclosed by the formed plume. The time selector color shows the degree values, reflecting the dynamic behavior of the plume towards the end. We chose representative time steps and scaled them to highlight the raising temperature; find the complete animation in the supplemental material.

\section{Conclusion and Future Work}\label{conclusion}
In this paper, we applied FCTs in the time-varying setting. As our new contribution, we adapted back-end and front-end to its specific challenges, resulting in the T-FCT interface. We illustrated usefulness and limitations in real-world examples. As future work, different options to deal with cluttered (Fuzzy) contour trees could be included, for example hiding nondescript branches and detail on demand. Also, instead of a purely additive comparison of contour trees, a subtractive comparison just showing branches not contained everywhere could improve the overview.
In summary, extending and applying FCTs to time varying data results in useful, coherent visualizations providing a more general overview of the data than common analysis approaches.

\acknowledgments{
Funded by the Deutsche Forschungsgemeinschaft (DFG, German Re-
search Foundation) – 252408385 – IRTG 2057}

\bibliographystyle{abbrv-doi}

\bibliography{references}
\end{document}